# The interplay between dopants and oxygen vacancies in the magnetism of V-doped TiO$_2$

**Ricardo Grau-Crespo[1*] and Udo Schwingenschlögl[2]**

[1]University College London, Department of Chemistry - 20 Gordon Street, WC1H 0AJ London, UK

[2]KAUST, PSE Division - Thuwal 23955-6900, Kingdom of Saudi Arabia

*Email: r.grau-crespo@ucl.ac.uk

**Abstract**

Density functional theory calculations indicate that the incorporation of V into Ti lattice positions of rutile TiO$_2$ leads to magnetic V$^{4+}$ species, but the extension and sign of the coupling between dopant moments confirm that ferromagnetic order cannot be reached via low-concentration doping in the non-defective oxide. Oxygen vacancies can introduce additional magnetic centres, and we show here that one of the effects of vanadium doping is to reduce the formation energies of these defects. In the presence of both V dopants and O vacancies all the spins tend to align with the same orientation. We conclude that V doping favours the ferromagnetic behaviour of TiO$_2$ not only by introducing spins associated with the dopant centres but also by increasing the concentration of oxygen vacancies with respect to the pure oxide.

Dilute magnetic oxides with Curie temperatures well above 300 K offer great promise for the development of spin injectors in spintronic devices that could operate at ambient conditions, among other exciting applications [1]. Thin films of various non-magnetic transition metal oxides, including $TiO_2$, $SnO_2$ and ZnO, have been found to exhibit room temperature ferromagnetism when doped with small concentrations of V, Cr, Co and other ions. In the case of anatase $TiO_2$, Hong et al. [2, 3] have indicated that vanadium is the most promising dopant, as it leads to a giant magnetic moment of 4.23 $\mu_B$ per dopant atom in the thin film, while Tian et al. [4] have recently observed ferromagnetism in both rutile and anatase $TiO_2$ thin films doped with 4 at.% vanadium. Although the nature of room temperature ferromagnetism in these oxides is not completely understood yet, most authors now agree that oxygen vacancies play an important role in the magnetic behaviour, at least in the case of $TiO_2$. For example, the ferromagnetism of V-doped $TiO_2$ samples is enhanced after annealing in an oxygen-deficient argon atmosphere, and the effect is more pronounced the higher the treatment temperature, which can be explained by the creation of oxygen vacancies during annealing [4]. Furthermore, it is known that even un-doped $TiO_2$ (both rutile and anatase) thin films can display ferromagnetic behaviour, but only if the preparation method favours the formation of oxygen vacancies (e.g. by vacuum annealing) [5-7]. Therefore, in order to understand the behaviour of dilute magnetic oxides it is essential to elucidate the interplay between dopants and defects.

We present here a theoretical investigation of the magnetism of V-doped $TiO_2$ (rutile) without and with oxygen vacancies. We simulate the incorporation of vanadium impurities using a 2x2x3 supercell of the original rutile unit cell, with none, one or two V/Ti substitutions, and with and without oxygen vacancies, leading to compositions $Ti_{24-n}V_nO_{48-m}$ ($n$=0,1,2; $m$=0,1). The number of calculations is significantly reduced by exploiting the symmetry of the system. For $n$=2 and $m$=0, for example, there are 24×23/2=276 possible cation distributions, but only 7 of them are symmetrically different, as determined with the SOD program [8]. Total energy, geometries and electronic structures were obtained using density functional theory (DFT) calculations in the generalised gradient approximation (GGA), with the exchange-correlation functional of Perdew, Burke and Ernzerhof [9], as implemented in the VASP program [10]. In order to improve the GGA description of the $d$ electrons in the solid, we used a Hubbard type correction for these orbitals following the so-called GGA+U approach, which acts by adding a positive term (proportional to a parameter $U_{eff}$) to the GGA energy, penalising the hybridisation of the specified orbitals of the metal atoms (e.g. Ti and V $d$ orbitals) with the ligands (*e.g.* O atoms) [11-13]. The introduction of the Hubbard correction (or an alternative correction to the self-interaction problem, *e.g.* a certain fraction of non-local exchange [14]) is necessary to reproduce the electronic, magnetic and redox properties of transition metal oxides, within the framework of local and semi-local DFT calculations (*e.g.* Refs. [15-17]). We have used $U_{eff}$=3 eV for both Ti and V 3$d$ orbitals, as this value has been shown to provide a good description of the electronic and redox properties of titanium and vanadium oxides [18, 19]. The interaction between the valence electrons and the core was described with the projected augmented wave (PAW) method in the implementation of Kresse and Joubert [20, 21]. The core electrons, up to 3$p$ in Ti and V and 1$s$ in O, were kept frozen in their atomic reference states. The cut-off energy controlling the number of plane wave basis functions was set to $E_{cut}$=520 eV, and all reciprocal space integrations were performed using a mesh of 2x2x2 points (for the 2x2x3 supercell). Ion positions and cell parameters were fully relaxed using a conjugate gradient algorithm. For all geometric configurations we explored the existence of different spin solutions by assigning initial local magnetic moments to the ions, which were then relaxed during the self-consistent field iterations, and also by fixing the total difference between up and down spins to the various allowed values.

The effect of an isolated V/Ti substitution on the electronic structure of $TiO_2$ is illustrated in Figure 1, which shows the calculated electronic density of states (DOS) for both pure rutile $TiO_2$ and V-doped rutile $Ti_{0.96}V_{0.04}O_2$ (one V dopant in the supercell). In pure rutile, the valence band is mainly of O 2$p$ character while the conduction band is mainly contributed by the Ti 3$d$ levels. The band gap of the pure oxide is underestimated in the calculations (~2 eV) in comparison with the experimental value (3.1 eV). Doping with V leads to an occupied $d$ level with majority spin at the top of the conduction band, and a more pronounced DOS peak with the same spin in the gap region, near the conduction band. The minority spin V 3$d$ levels are all empty. This electronic configuration corresponds to magnetic $V^{4+}$ species ($3d^1$).

Considering the case of two dopants in the simulation cell allows us to investigate the strength and character of dopant-dopant interactions. Table 1 shows some of the calculated properties for the 7 different configurations with cell composition $Ti_{22}V_2O_{48}$. It is clear that the energy changes considerably with the relative position of the V ions. The dopant-dopant interaction energy,

$$E_{VV} = E[Ti_{22}V_2O_{48}] + E[Ti_{24}O_{48}] - 2E[Ti_{23}VO_{48}] \qquad (1)$$

(the energy of the cell with two V substitutions is taken here as the average between the ferromagnetic and the antiferromagnetic configurations), is negative, suggesting a trend of the V ions to clusterize. The absolute value of the interaction energy decreases with the separation between the V centres, as shown in Fig. 2. The two lowest-energy configurations (most negative interaction energies) are those in which the V ions are in nearest neighbour (NN) positions. Two types of NN positions can occur, which are represented in the inset of Fig. 2: one with $VO_6$ octahedra sharing edges and another with $VO_6$ octahedra sharing corners. The former is the most stable, but there are more pairs of the second type in the lattice. It is possible to estimate the probabilities of occurrence for the different configurations using canonical statistical mechanics [8, 22, 23]:

$$P_m = \frac{\Omega_m}{Z} \exp(-E_m / k_B T),  \tag{2}$$

where $m=1, \ldots, 7$ is the index of the configuration, $E_m$ is the energy of the configuration, $\Omega_m$ is its degeneracy (the number of times that configuration $m$ is repeated in the complete configurational space) and $k_B$ is Boltzmann's constant. Table 1 shows that at 1000 K around half of the V ions at this concentration are expected to be forming NN pairs, with 40% in corner-sharing octahedra and 10% in edge-sharing octahedra. Notably, the two most stable configurations, forming NN pairs, exhibit ferromagnetic coupling, while V-V pairs at longer distances are coupled antiferromagnetically. This contrasts with the ferromagnetic coupling at long distances (first to fifth neighbours) found for anatase $TiO_2$ by Osorio-Guillen *et al.,* using similar DFT calculations [24]. Obviously, in the case of rutile, the NN coupling cannot explain the long-range ferromagnetism reported in experiment for low V concentrations. At most these results imply that the presence of isolated pairs (or maybe small clusters) of vanadium dopants with parallel spin alignment is thermodynamically favourable. However, it is very well possible that the distribution of vanadium in real samples is not in configurational equilibrium, and is instead randomized by kinetic factors during growth, which would lead to a more homogeneous distribution of the dopants.

In order to examine the role of oxygen vacancies in the magnetic behaviour of V-doped $TiO_2$, we have calculated the energies of all different configurations with one V substitution and one O vacancy in the supercell at their equilibrium geometries, which allows us to obtain the vacancy formation energies:

$$\text{VFE} = E[\text{VTi}_{23}\text{O}_{47}] + \frac{1}{2} E[O_2] - E[\text{VTi}_{23}\text{O}_{48}]. \tag{3}$$

Since two excess electrons are introduced by the oxygen vacancy and one by the vanadium dopant, we calculate solutions where the total magnetization per cell is $M=1$ or 3. For comparison we also calculate the formation energy of a vacancy in pure $TiO_2$ with both parallel and antiparallel arrangements of the spins of the excess electrons ($M=0$ or 2, respectively). The calculated values are shown in Figure 3. For pure $TiO_2$, the groundstate vacancy configuration corresponds to a solution where the two excess electrons are oriented in the same direction, although the antiparallel spin configuration is only 20 meV higher in energy. The value of the vacancy formation energy (VFE=4.78 eV) is in reasonable agreement with previous estimations (e.g. 4.18 eV in Ref. [25]). According to our calculations, the excess electrons associated to the vacancy distribute on the neighbouring Ti cations, although not only on the nearest neighbours. In particular, we do not observe any electron localisation at the vacancy site, as suggested by earlier Hartree-Fock calculations [26].

The formation energies of oxygen vacancies decrease significantly in the presence of a vanadium dopant, as shown in Figure 3. Interestingly, this occurs for almost all relative positions of the vacancy with respect to the dopant in the cell (there are 12 symmetrically different configurations for this supercell) and not only when they are nearest neighbours. It can also be observed that, in general, high-spin solutions with parallel spin orientations ($M=3$) have lower energies than those with $M=1$. The decomposition of the charge density indicates that, in all the high-spin solutions, one of the excess electrons associated with the vacancy localises on the vanadium dopant, forming a high-spin $V^{3+}$ species, while the other excess electron localises over one or two Ti sites. In particular, in the lowest energy $VTi_{23}O_{47}$ configuration there is a $V^{3+}$ and a well-defined $Ti^{3+}$ centre, but none of these reduced cations is nearest neighbour to the vacancy, as would be expected based on a simple electrostatic picture (the vacancy defect is positive while the 3+ centres are negatively charged with respect to the original ions). This behaviour arises when local relaxation effects dominate over electrostatic factors, and has been reported in other cases, for example the localization of positive oxygen vacancies in ceria and in Ti-doped zirconia surfaces [27, 28].

The described behaviour suggests that V dopants can contribute to the ferromagnetism of titanium oxide not only by providing magnetic centres associated to their *d* electrons, but also by increasing the concentration of oxygen vacancies with respect to the pure oxide. It could be argued that the oxygen vacancy formation energies are still too high, even after the stabilisation provided by the vanadium dopants, to give a significant concentration of vacancies. However, the values given here should still be corrected for vibrational effects. Keith et al. [25] have recently estimated, using ab initio molecular dynamics and thermodynamic integration techniques, that the formation free energy of an oxygen vacancy in pure $TiO_2$ decreases by 2.3 eV when increasing the temperature from 0 K to 266 K. It is very difficult to produce estimates of the concentrations of vacancies under different conditions of doping, oxygen partial pressure and temperature, as this would require a statistical treatment incorporating entropic effects of configurational, vibrational and magnetic origin, and still would be very sensitive to the error involved in the DFT calculation of energies. Nevertheless, the present results clearly indicate that the formation of vacancies in $TiO_2$ is strongly favoured by the presence of vanadium dopants, which brings new insight into the role of dopants as promoters of magnetic behaviour in oxides.

This work made use of the facilities of HECToR, the UK's national high-performance computing service, via RGC's membership of the UK's HPC Materials Chemistry Consortium, which is funded by EPSRC (EP/F067496).

**Table 1.**

List of configurations with supercell composition $Ti_{22}V_2O_{48}$. $d_{VV}$ is the shortest V-V distance in the supercell, $\Omega$ is the configuration degeneracy, $E_{VV}$ is the dopant-dopant interaction energy, $P_{1000K}$ is the probability of occurrence of the configuration upon equilibration at 1000 K, and $\Delta E_{FM-AF}$ is the difference in energy between the ferromagnetic (FM) and the antiferromagnetic (AF) orientation of the V spin moments.

| $d_{VV}$ (Å) | $\Omega$ | $E_{VV}$ (meV) | $P_{1000K}$ | $\Delta E_{FM-AF}$ (meV) |
|---|---|---|---|---|
| 3.01 | 24 | -151 | 0.109 | -7.6 |
| 3.60 | 96 | -142 | 0.394 | -5.6 |
| 4.68 | 24 | -118 | 0.075 | 52.7 |
| 5.57 | 48 | -113 | 0.141 | 55.1 |
| 5.60 | 48 | -138 | 0.188 | 1.5 |
| 6.61 | 12 | -81 | 0.024 | 4.2 |
| 7.26 | 24 | -110 | 0.068 | 1.1 |

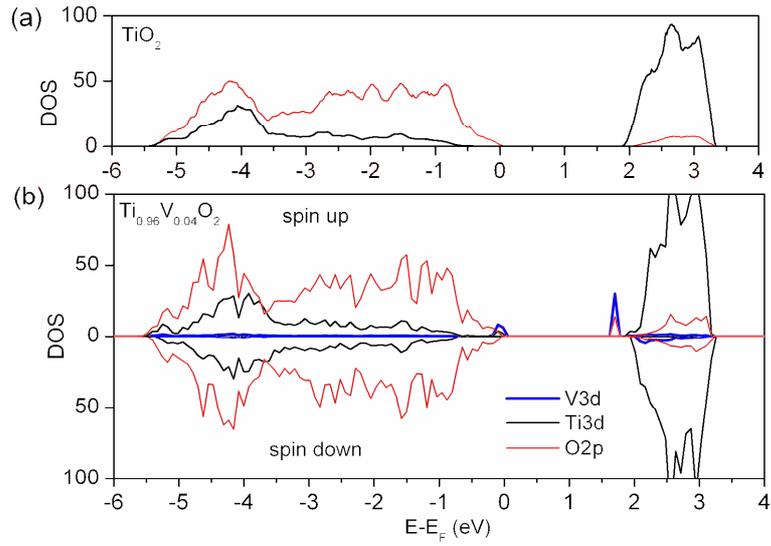

**Figure 1.**

Calculated density of states for a) pure $TiO_2$ and b) $TiO_2$ doped with isolated V ions at 4 at.% concentration.

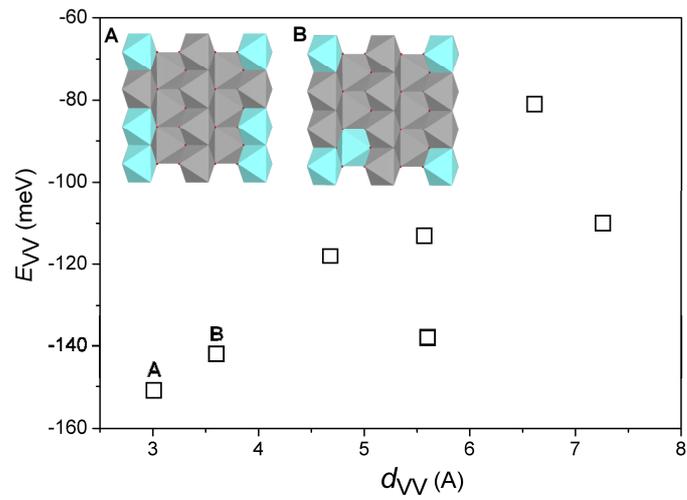

**Figure 2.**
**Dopant-dopant interaction plotted against the shortest V-V distance in the supercell. The inset shows the two different nearest-neighbour configurations, with edge-sharing (A) and corner-sharing (B) $VO_6$ octahedra. Gray and light blue octahedra are centred by Ti and V cations, respectively.**

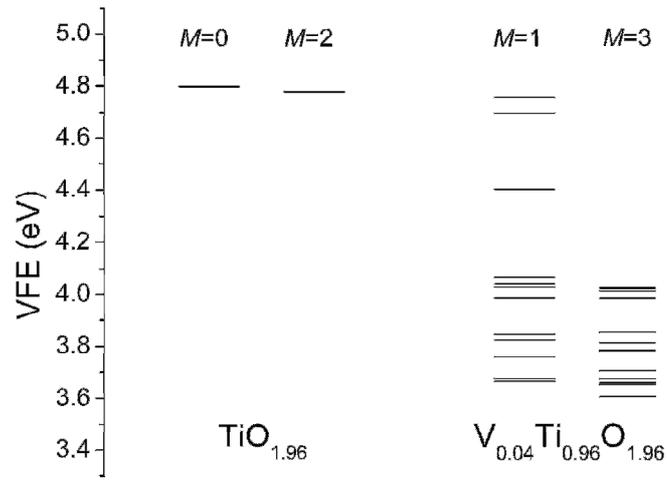

**Figure 3.**

**Oxygen vacancy formation energies (VFE) for pure and vanadium-doped titanium oxide. *M* is the total magnetization of the cell.**